\newif\ifreview
\reviewfalse        

\ifreview
  \documentclass[preprint,1p,times]{elsarticle}
\else
  \documentclass[5p,times]{elsarticle}
\fi

\ifreview
  \usepackage{lineno}
\fi
\usepackage{graphicx}
\usepackage{amsmath}
\usepackage{amssymb}
\usepackage{booktabs}
\usepackage{array}
\usepackage{multirow}
\usepackage{subcaption}
\usepackage{siunitx}
\usepackage{xcolor}
\usepackage{float}

\journal{Nuclear Instruments and Methods in Physics Research A}

\begin{document}

\ifreview
\linenumbers
\fi

\begin{frontmatter}

\title{Two Stage Gamma--Neutron Source Classification in Water Cherenkov Detectors:
Energy Threshold Screening and Machine Learning Pulse Analysis}

\author[1,2]{Alejandro Núñez Selin\corref{cor1}}
\author[1]{Iván Sidelnik}
\author[3,4]{Christian Sarmiento Cano}
\author[5]{Hernán Asorey}
\author[3,6]{Luis A. Núñez}

\address[1]{Departamento de Física de Neutrones, Centro Atómico Bariloche (CNEA/CONICET),
Av. Bustillo 9500, S. C. de Bariloche, Argentina}
\address[2]{Instituto Balseiro, CNEA, Av. Bustillo 9500, San Carlos de Bariloche, Argentina}
\address[3]{Departamento de Ciencias Básicas, Universidad Autónoma de Bucaramanga, Bucaramanga, Colombia}
\address[4]{Escuela de Física, Universidad Industrial de Santander, Bucaramanga, Colombia}
\address[5]{piensas.xyz, Las Rozas Innova, Jacinto Benavente 2, 28232 Las Rozas de Madrid, Spain}
\address[6]{Depatamento de F\'isica, Universidad de Los Andes, M\'erida, Venezuela}

\begin{abstract}
Water Cherenkov detectors offer a robust and economical solution for real time radiation monitoring by detecting Cherenkov light from charged particles moving faster than light in water. This work presents a novel two stage classification framework for gamma--neutron discrimination: an initial physics-based energy threshold filters unambiguous low energy gamma sources, followed by a machine learning ensemble that resolves ambiguities at higher energies. The detector response was characterized using $^{60}$Co (1.17/1.33~MeV), $^{137}$Cs (0.66~MeV), and a shielded $^{241}$AmBe source, with lead, paraffin, and cadmium shielding employed to isolate neutron and gamma interactions. Energy calibration established a linear ADU to MeV conversion ($R^2 = 0.966$), enabling identification of a neutron detection threshold at $2.62 \pm 0.77$~MeV via a $3\sigma$ significance analysis. Stage one categorizes sources as pure gamma (below threshold) or neutron emitting (at threshold). For ambiguous cases above threshold, a machine learning pipeline utilizing pulse shape analysis was developed. A soft voting ensemble (Bagging, CatBoost, and MLP) achieved an accuracy of 0.816 and an AUC of 0.921. This hybrid scheme combines physics based filtering with ML refinement, offering an interpretable and scalable solution for nuclear security, nonproliferation monitoring, and fundamental radiation research.
\end{abstract}


\end{frontmatter}

\section{Introduction}

Water Cherenkov Detectors (WCDs) exploit the Cherenkov effect to detect relativistic charged particles, finding applications in fundamental physics \cite{fukuda2003super,auger_observatory_2015} and nuclear security \cite{sidelnik2020enhancing}. The growing scarcity of traditional \textsuperscript{3}He detectors \cite{sachetti_3he_free_2015} has intensified interest in water-based alternatives for neutron-gamma discrimination. Pure water WCDs offer inherent advantages: scalability, non-toxicity, and the ability to leverage temporal pulse characteristics for particle identification \cite{watanabe_neutron_tagging_2009}.

This work addresses two critical challenges in operationalizing water based detectors for security applications: (1) establishing reliable energy deposition thresholds for coarse particle-type discrimination, and (2) developing pulse level classification methods for mixed radiation fields. Our experimental framework employs pure gamma emitters, \textsuperscript{60}Co (1.17/1.33 MeV) and \textsuperscript{137}Cs (0.66 MeV), and a mixed \textsuperscript{241}AmBe neutron(0--11 MeV)--gamma(4.44 MeV) source, under controlled shielding configurations. Lead, paraffin and Cadmium shielding configurations enable isolation of neutron and gamma components through differential attenuation properties \cite{chichester2007radiation}. We first develop a calibration protocol linking deposited energy to incident radiation characteristics, establishing statistically significant thresholds for neutron detection. Subsequently, we implement an ensemble machine learning classifier optimized for pulse level discrimination in ambiguous energy regimes. 

This integrated methodology maintains water's inherent advantages while overcoming traditional limitations in signal discrimination, directly addressing operational requirements in nuclear security applications where reliable classification must coexist with system interpretability.

\section{Experimental Setup}

\subsection{Detector System}
The WCD at the Bariloche Atomic Center Neutron Physics Laboratory features a 1.0 m$^3$ cylindrical volume (94 cm diameter $\times$ 147 cm height) filled with pure water, serving as both Cherenkov radiator and neutron moderator. Photodetection is provided by a Photonis XP1802 9'' photomultiplier tube (PMT) operated at 1230 V, with spectral response optimized for Cherenkov emission (300--650 nm). Signals are digitized by a Red Pitaya STEMLAB 125-14 FPGA sampling at 125 MS/s in ADU (Analogic to Digital units) within the LAGO DAQ ecosystem \cite{LAGO_rep}. Interior surfaces are lined with Tyvek\textregistered{} for enhanced diffuse reflection, maximizing photon collection efficiency.

\subsection{Radiation Interactions and Detection Principles}
In pure water ($n = 1.33$), the electron Cherenkov threshold is 262 keV. Gamma radiation interacts primarily through Compton scattering, photoelectric effect, or pair production (for $E_\gamma > 1.022$ MeV), generating secondary electrons that emit Cherenkov photons. Neutrons undergo moderation via elastic scattering with hydrogen nuclei, followed by radiative capture ($n + ^1\text{H} \rightarrow ^2\text{H} + \gamma$, $E_\gamma = 2.22$ MeV). This characteristic 2.22 MeV gamma serves as the primary neutron signature in water-based detection.

\subsection{Shielding Configurations}

Strategic shielding configurations isolated specific radiation components from the \textsuperscript{241}AmBe source:
\begin{itemize}
    \item \textbf{Pb (10 cm):} Attenuates 4.44 MeV gammas, isolating the neutron spectrum
    \item \textbf{Borated Paraffin (10 cm) + Cd (0.1 cm):} Absorbs neutrons, selecting the 4.44 MeV gamma component
    \item \textbf{Pb (10 cm) + Pure Paraffin (15 cm):} Attenuates 4.44 MeV gammas, then produces external 2.22 MeV gammas via neutron capture in paraffin.
\end{itemize}

\section{Energy-Level Classification}

\subsection{Cutoff Point Determination}
The cutoff point is defined as the energy at which the \emph{histogrammed source counts} become statistically indistinguishable from the background, according to a $3\sigma$ criterion. For each histogram bin, the combined uncertainty is computed as:
\begin{equation}
\sigma = \sqrt{\sigma_{\mathrm{src}}^2 + \sigma_{\mathrm{bkg}}^2}, \quad \text{with} \quad \sigma_{\mathrm{src}} = \sqrt{N_{\mathrm{src}}}, \quad \sigma_{\mathrm{bkg}} = \sqrt{N_{\mathrm{bkg}}},
\label{eq:sigma_total}
\end{equation}
where \(N_{\mathrm{src}}\) and \(N_{\mathrm{bkg}}\) are the counts from the source and background, respectively. The cutoff is defined as the first bin for which:
\begin{equation}
\left|N_{\mathrm{src}} - N_{\mathrm{bkg}}\right| < 3\sigma.
\label{eq:3sigma}
\end{equation}

The cutoff uncertainty $\Delta_{\text{cutoff}}$ combines \textbf{statistical}: bootstrap resampling (1000 iterations)\cite{davison1997bootstrap}, \textbf{systematic}: semi-range between $1\sigma$ and $5\sigma$ cutoffs, capturing the sensitivity to the threshold choice and \textbf{discretization}: half bin width, arises from the finite bin width. These contributions are combined in quadrature:

\begin{equation}
\Delta_{\mathrm{cutoff}} = \sqrt{\Delta_{\mathrm{stat}}^2 + \Delta_{\mathrm{sys}}^2 + \left(\frac{\Delta_{\mathrm{bin}}}{2}\right)^2}.
\label{eq:delta_cutoff}
\end{equation}

\subsection{Energy Calibration}

\begin{table}[H]
\centering
\begin{tabular}{lcc}
\toprule
\textbf{Source (Configuration)} & \textbf{Energy (MeV)} & \textbf{Cutoff (ADU)} \\
\midrule
\textsuperscript{137}Cs & 0.66 & $1090 \pm 83$ \\
\textsuperscript{60}Co & 1.17, 1.33 & $2130 \pm 175$ \\
\textsuperscript{241}AmBe (Pb) & 2.22 & $9010 \pm 1271$ \\
\textsuperscript{241}AmBe (Pb/Paraffin) & 2.22 & $8670 \pm 1171$ \\
\textsuperscript{241}AmBe (B-Paraffin/Cd) & 4.44 & $14890 \pm 2967$ \\
\textsuperscript{241}AmBe (Unshielded) & 2.22, 4.44 & $16490 \pm 4480$ \\
\bottomrule
\end{tabular}
\caption{Cutoff points for different sources and configurations.}
\label{tab:cutoffs}
\end{table}

\begin{figure}[h!]
\centering
\begin{subfigure}{0.48\linewidth}
\includegraphics[width=\linewidth]{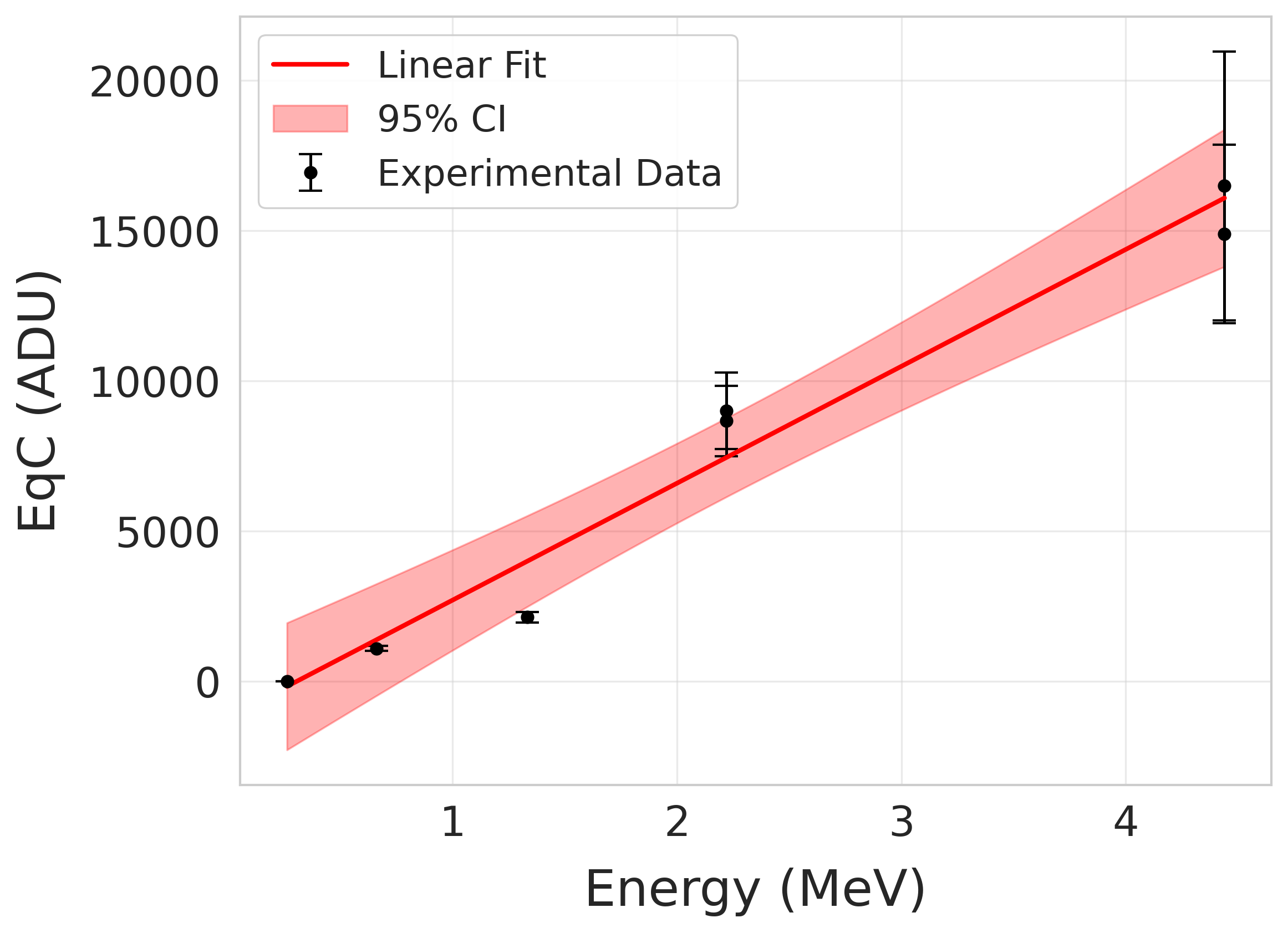}
\caption{Calibration curve with 95\% confidence band.}
\label{fig:calibration}
\end{subfigure}
\hfill
\begin{subfigure}{0.48\linewidth}
\includegraphics[width=\linewidth]{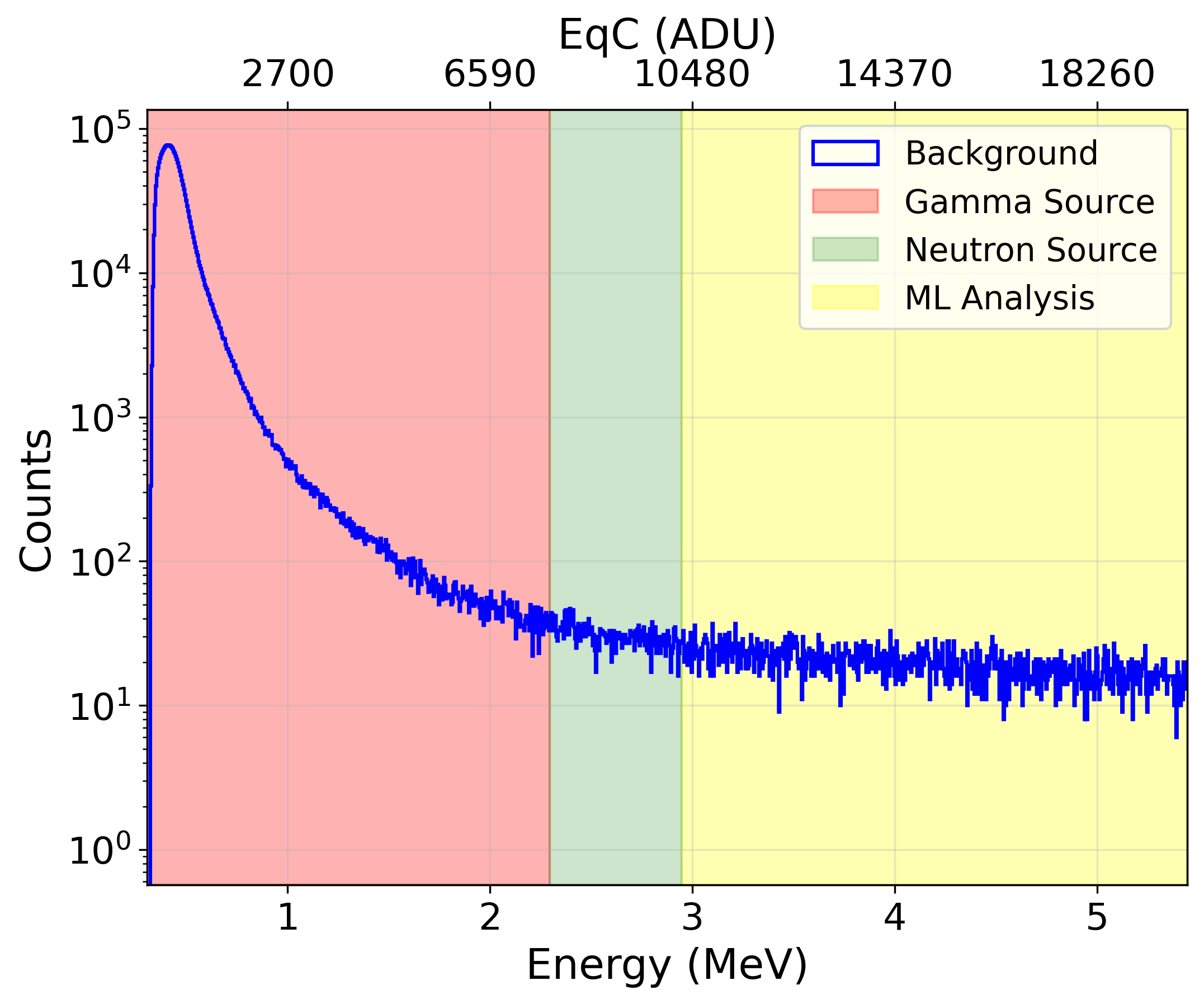}
\caption{Converted energy spectrum}
\label{fig:adu-energy}
\end{subfigure}
\caption{(a) ADU to Energy conversion.  The background-equivalent charge (ADUs) is converted into radiation energy (MeV) using Eq.~\ref{eq:fit}. (b) The green band represents the neutron detection threshold, marking the transition from red (no neutrons) to yellow (possible neutron presence).}
\end{figure}

Table~\ref{tab:cutoffs} reports the resulting cutoff points. These ADU values, when associated with known gamma energies (including the 2.22 MeV neutron capture line), establish a linear charge to energy calibration, incorporating the electron Cherenkov threshold (0.262 MeV), Fig \ref{fig:calibration}.
\begin{equation}
\text{EqC(ADU)} = (3.89 \pm 0.84)\times10^3 \, E(\text{MeV}) - (1.19 \pm 2.29)\times10^3,
\label{eq:fit}
\end{equation}

with $R^2 = 0.966$. The calibrated cutoff for the neutron-induced 2.22 MeV gamma defines the neutron detection threshold at $2.62 \pm 0.77$ MeV. Radioactive sources below this threshold are classified as unambiguous gamma-only source, Fig \ref{fig:adu-energy}.

\section{Pulse-Level Classification}

\subsection{Data Preprocessing}
Pulses from \textsuperscript{241}AmBe (Pb and Cd/B-paraffin configurations) in the 4500--8000 ADU range were used. Each pulse has 32 temporal bins (8 ns each). The dataset was split 80/20 (stratified), with Random Under Sampling (RUS) \cite{lemaavztre2017imbalanced} applied to the training set to balance classes (82,967 neutron and gamma pulses each). Features were standardized using \texttt{StandardScaler} method from \texttt{scikit-learn} ~\cite{pedregosa2011scikit}. 

\subsection{Machine Learning Pipeline}
To develop a robust classifier for ambiguous high energy pulses, we evaluated over 25 algorithms spanning diverse families including tree based ensembles, neural networks, support vector machines, and distance-based models. From these, 12 classifiers exceeded our accuracy threshold of 0.75. To leverage complementary strengths, we constructed a soft voting ensemble \cite{kuncheva2014combining} combining the three best-performing classifiers from different algorithm families: Bagging \cite{breiman1996bagging}, CatBoost \cite{prokhorenkova2018catboost}, and MLP \cite{goodfellow2016deep}. This selection was based on both performance metrics and prediction diversity, ensuring the ensemble benefits from decorrelated errors. Hyperparameter optimization was performed via \texttt{RandomizedSearchCV} (100 iterations, 5-fold CV) \cite{bergstra2012random}.

Predictions are combined via optimized soft voting, with weights assigned proportionally to individual F1-scores. Threshold optimization balanced true positive and negative rates, yielding an optimal decision threshold of 0.52.

The optimized ensemble achieved strong performance on the test set: accuracy = $0.816$ (95\% CI: 0.814--0.818), AUC = $0.921$ (95\% CI: 0.919--0.922), with balanced F1-scores ($\sim$0.815) for both classes, confirming unbiased discrimination. Detailed performance diagnostics (e.g., confusion matrices and ROC curves) are omitted
for brevity and will be addressed in a forthcoming extended journal publication.

\begin{figure}[H]
    \centering
        \includegraphics[width=0.7\linewidth]{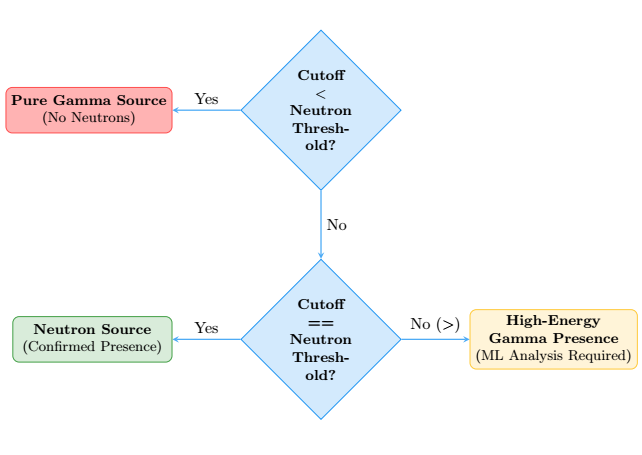}
        \caption{Classification workflow: Decision nodes (blue) route sources to pure gamma (red), confirmed neutron (green), or ambiguous cases (yellow). Machine learning refines high-energy ambiguities to final neutron/gamma states.}      
    \label{fig:flux_diagram}
\end{figure}

\section{Conclusions}

We developed and validated a two stage gamma--neutron discrimination framework for water Cherenkov detectors that integrates a physics based statistical analysis with machine learning pulse classification.

In the first stage, a robust energy calibration using $^{60}$Co, $^{137}$Cs, and multiple shielding configurations of a $^{241}$AmBe source established a linear ADU--MeV relationship ($R^2 = 0.966$). A $3\sigma$ significance analysis identified a neutron detection threshold at $2.62 \pm 0.77$ MeV, enabling rapid and interpretable classification of pure gamma sources and confirmed neutron emitting fields.

For events above this threshold, a second stage soft voting ensemble combining Bagging, CatBoost, and MLP classifiers achieved reliable pulse-level discrimination, reaching 0.816 accuracy and 0.921 AUC. The ensemble leverages complementary learning strategies while maintaining moderate inter model correlation, ensuring non redundant predictive information. Threshold optimization further improved the balance between sensitivity and specificity.

The combined framework operates as an effective “traffic light” system: below threshold events are classified as gamma only, threshold level events confirm neutron presence, and above-threshold cases are resolved through machine learning, Fig \ref{fig:flux_diagram}. This approach preserves the intrinsic advantages of water based detectors scalability, safety, and low cost while addressing key limitations in neutron--gamma separation.

Overall, the proposed methodology provides an interpretable, scalable, and operationally efficient solution for neutron detection in nuclear security and nonproliferation contexts. Future work will focus on deep learning architectures for raw waveform analysis and improved performance in low energy, background dominated regimes.

\section*{Acknowledgements}

The authors acknowledge co-funding from the Programa Iberoamericano de Ciencia y
Tecnología para el Desarrollo (CYTED) through the LAGO-INDICA network
(Project 524RT0159-LAGO-INDICA: Infraestructura digital de ciencia abierta). This work was partially supported by the CLAF-HECAP Programme through a research
mobility scholarship.

\end{document}